\documentclass{aa}  
\usepackage{graphicx}
\usepackage{txfonts}

\begin{document} 

\title{Flows and magnetic field structures in reconnection regions of simulations of the solar atmosphere: do flux pile-up models work?}

\titlerunning{Flows and magnetic fields in solar photospheric reconnection regions}
\authorrunning{Shelyag et al.}

\author{S. Shelyag\inst{1}, Y. E. Litvinenko\inst{2}, V. Fedun\inst{3}, G. Verth\inst{4}, J.~J. Gonz\'alez-Avil\'es\inst{5,6}, F.~S. Guzm\'an\inst{6}} 
\institute{Department of Mathematics, Physics and Electrical Engineering, Northumbria University, Newcastle upon Tyne, NE1 8ST, UK \and
Department of Mathematics, University of Waikato, P. B. 3105, Hamilton, New Zealand \and
Plasma Dynamics Group, Department of Automatic Control and Systems Engineering, University of Sheffield, Sheffield, S1 3JD, UK \and
Plasma Dynamics Group, School of Mathematics and Statistics, University of Sheffield, Sheffield, S3 7RH, UK \and
Instituto de Geof\'isica, Unidad Michoac\'an, Universidad Nacional Aut\'onoma de M\'exico, Morelia, Michoac\'an, M\'exico \and
Laboratorio de Inteligencia Artificial y Superc\'omputo. Instituto de F\'isica y Matem\'aticas, Universidad Michoacana de San Nicol\'as de Hidalgo. Morelia, Michoac\'an, M\'exico}

\date{Received; accepted}

  \abstract
{}
{We study the process of magnetic field annihilation and reconnection in simulations of magnetised solar photosphere and chromosphere with magnetic fields of opposite polarities and constant numerical resistivity.}
{Exact analytical solutions for reconnective annihilations are used to interpret the features of magnetic reconnection in simulations of flux cancellation in the solar atmosphere. We use MURaM high-resolution photospheric radiative magneto-convection simulations to demonstrate the presence of magnetic field reconnection consistent with the magnetic flux pile-up models. Also, a simulated data-driven chromospheric magneto-hydrodynamic simulation is used to demonstrate magnetic field and flow structures, which are similar to the ones theoretically predicted.}
{Both simulations demonstrate flow and magnetic field structures roughly consistent with accelerated reconnection with magnetic flux pile-up. The presence of standard Sweet-Parker type reconnection is also demonstrated in stronger photospheric magnetic fields.}
{}


\maketitle

%

\section*{Introduction}

The observational term ``cancellation'' describes the disappearance of magnetic flux of either sign at the polarity inversion line that separates the magnetic fragments with opposite polarity in the solar photosphere \citep{livi-et-al-1985,martin-et-al-1985}. Photospheric cancellation appears to be a key dynamic process in the removal of solar magnetic flux and in the formation and evolution of solar filaments \citep{martens-zwaan-2001,martin-et-al-2008,panasenco-et-al-2014}. Cancellation remains a subject of active research, based on the data from several instruments, including those on {\it Solar Dynamics Observatory} \citep{zeng-et-al-2013,2016ApJ...827..151Y}.

Observations of evolving magnetic features in the photosphere strongly suggest that magnetic reconnection in a photospheric or chromospheric current sheet, rather than simple submergence, is the cancellation mechanism \citep{martin-1990,chae-2012}. Photospheric magnetic fragments originate as bipoles but cancel with external fields. Fragments with the same polarity do not cancel on encounter but rather merge to form a single larger magnetic feature, whereas cancelling fragments of opposite polarity usually slow down on encounter, indicating that mutual interaction takes place. On the theoretical side, a model of flux pile-up reconnection in a Sweet--Parker current sheet \citep{1957JGR....62..509P}, suitably modified for a compressible, weakly ionized photospheric plasma, can explain the properties of cancelling magnetic features, such as the speeds of the cancelling magnetic fragments and the flux cancellation rates, inferred from the data \citep{litvinenko-1999,litvinenko-et-al-2007,park-et-al-2009}.

Magnetic energy release in a chromospheric reconnecting current sheet leads to bulk heating of the chromospheric plasma, balanced by radiative cooling \citep{litvinenko-somov-1994}. Except for very small cancelling features, thermal conduction can be neglected. Radiation is the dominant mechanism of energy loss from the chromospheric current sheet. The high density and low temperature in the current sheet also mean that particle acceleration by the reconnection electric field is inefficient \citep{2015JKAS...48..187L}. 

Reconnection also converts a part of the free magnetic energy into the kinetic energy of reconnection jets. The jets travel with a local Alfv\'en speed $v_A$ which is of order a few km/s in the photosphere. H$\alpha$ and magnetogram data indeed show that photospheric cancellation is accompanied by plasma upflows \citep{1999SoPh..190...45L,2005ApJ...626L.125B} and downflows \citep{2004ApJ...602L..65C}. The speeds of ubiquitous quiet-Sun jets \citep{2011A&A...530A.111M} are consistent with the reconnection outflow speeds in the range of $3-10$ km s$^{-1}$, predicted by photospheric reconnection models \citep{litvinenko-1999,litvinenko-et-al-2007}.

Much faster chromospheric jets are associated with explosive events, detected by ultraviolet (UV) and extreme-ultraviolet (EUV) observations of the upper chromosphere and transition region. The explosive events correlate with photospheric cancellation \citep{1991JGR....96.9399D} or, more generally, with changes of the photospheric magnetic structure \citep{2008ApJ...687.1398M}. The photospheric jets with speeds of the order of a few km s$^{-1}$ and the chromospheric jets with speeds of up to 100 km s$^{-1}$ are thought to be direct signatures of magnetic reconnection at the corresponding heights \citep{1997Natur.386..811I,2016MNRAS.463.2190N}. High-resolution observations of Ellerman bombs \citep{2011ApJ...736...71W,2013ApJ...774...32V} suggest that local photospheric reconnection can cause the magnetic field relaxation on a larger scale, leading to the photospheric and chromospheric jet generation \citep{reid-et-al-2015}. Additionally, \citet{2015ApJ...811...48Y} argued that self-absorption features in transition region lines imply similar magnetic field changes in a range of observational phenomena (explosive events, blinkers, Ellerman bombs), which only differ by the height of a magnetic reconnection site. It appears, therefore, that photospheric cancellation can create favourable conditions for the generation of the photospheric and chromospheric jets, either directly or by triggering the release of stored magnetic energy on a larger scale. 

In this paper, we use a combination of numerical magneto-hydrodynamic simulations and exact analytical solutions for magnetic field reconnection to demonstrate the presence of magnetic pile-up mechanism, accelerating reconnection process. Also, we demonstrate the presence of outflows produced by the reconnection with magnetic pile-up. 

The paper is organised as follows. In Section 1, we outline an analytical model for reconnection with magnetic field pile-up. Section 2 describes the simulation setup we use for the analysis of reconnection regions in photospheric magneto-convection simulations. In Section 3, using averaging over the current sheet surroundings in the simulated reconnection regions, we demonstrate presence of magnetic pile-up, consistent with the theoretical model. Section 4 is devoted to the simulated chromospheric reconnection and flow structure in the reconnection region. Section 5 concludes our findings.

\section{Analytical models of magnetic annihilation with flux pile-up}

Magnetic flux pileup merging (annihilation) is one of the few models of magnetic reconnection for which detailed analytical description is available \citep{Priest-Forbes-2000}. Exact analytical solutions for the annihilation of planar magnetic fields, driven by a stagnation-point flow in an incompressible resistive plasma, were discovered by \citet{1964PhFl....7.1299C,1965PhFl....8..644C} and independently by \citet{1973JPlPh...9...49P} and \citet{1975JPlPh..14..283S}. Later the solutions were generalised to describe reconnective annihilation of magnetic fields in a current sheet in two and three dimensions  \citep{1995ApJ...450..280C,1995ApJ...455L.197C,1996ApJ...462..969C} and incorporate numerous potentially important effects, such as plasma viscosity \citep{2012ApJ...747...16C} and a non-vanishing curvature of the current sheet \citep{2002SoPh..207..337W,2013ApJ...774..155L}. Although no analogous exact solutions for magnetic merging in a compressible plasma had been obtained, the incompressible reconnection model with flux pileup was argued to yield robust magnetic reconnection scalings \citep{2003SoPh..218..173L}. 

Predictions of the analytical theory of flux pileup merging were found to be in good quantitative agreement with the results of numerical simulations performed in a two-dimensional periodic geometry  \citep{2000GApFD..93..115H}. Yet we are not aware of a detailed application of the theory to simulation results obtained in a more realistic geometry lacking a high symmetry, which motivates us to employ the analytical model to investigate how the geometry of magnetic merging controls the observable signatures of magnetic reconnection in our simulations of the photospheric dynamics. Specifically we are interested in the role played by the velocity and magnetic field geometry in quantifying the rate of magnetic energy dissipation and the generation of vertical reconnection-related jets. Following \citet{1995ApJ...455L.197C}, we present an illustrative solution in the limiting case of the planar annihilating magnetic field lines that is parallel to the plane of the velocity field of a stagnation-point flow. 

\begin{figure}
\centering
\includegraphics[width=\linewidth]{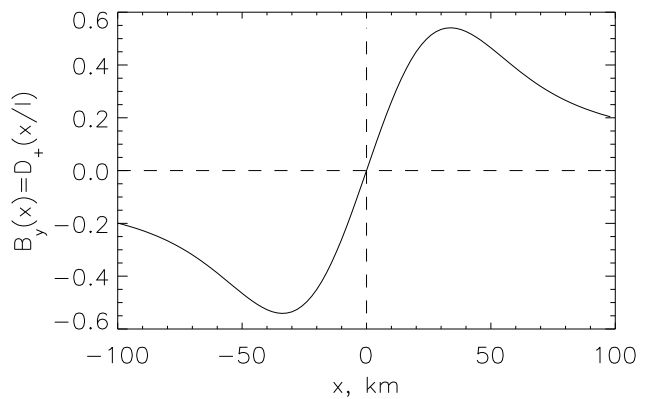}
\caption{Dawson function, corresponding to the reconnecting magnetic field profile around the current sheet. The current sheet thickness $l$ is chosen according to the simulation parameters.}
\label{fig1_dawson}
\end{figure}

We seek solutions of the governing magnetohydrodynamic (MHD) equations for the velocity ${\bf v}$ and magnetic field ${\bf B}$ in an incompressible resistive plasma: the momentum equation 
\begin{equation} 
  \partial_t {\bf v} + ({\bf v} \cdot \nabla) {\bf v} = -\frac{\nabla p}{\rho} + \frac{1}{\rho}(\nabla \times {\bf B}) \times {\bf B}, 
\end{equation}
the induction equation 
\begin{equation} 
  \partial_t {\bf B} = \nabla \times ({\bf v} \times {\bf B}) + \eta \nabla^2 {\bf B} , 
  \label{inductioneq}
\end{equation}
the incompressible continuity equation 
\begin{equation} 
  \nabla \cdot {\bf v} = 0 , 
\end{equation}
and the divergence-free condition for the magnetic field
\begin{equation} 
  \nabla \cdot {\bf B} = 0 . 
\end{equation}

Here, $\rho$ is the plasma density, $\eta$ is the magnetic diffusivity, and $p$ is the plasma pressure. The magnetic field in the equations is normalised by $\sqrt{4\pi}$ for convenience.

\begin{figure*}
\centering
\includegraphics{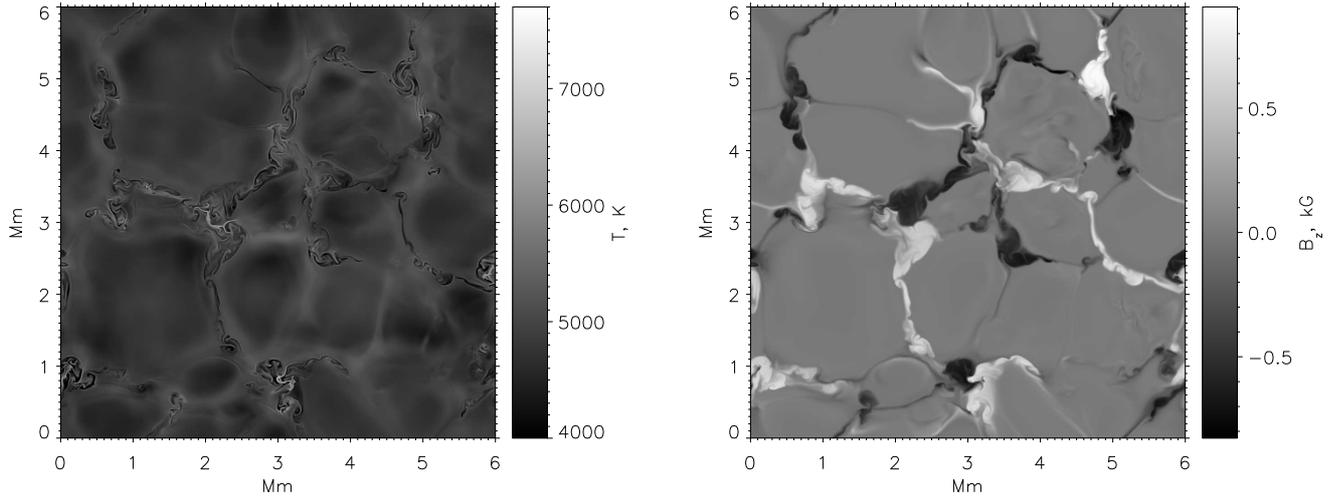}
\caption{Horisontal cuts of the temperature (left panel) and the vertical component of magnetic field (right panel) in the domain, taken at a height of $\sim 300~\mathrm{km}$ above the continuum formation height. Small-scale temperature enhancements up to $7700~\mathrm{K}$ indicate current sheets in the centres of magnetic reconnection regions. These regions can be clearly seen in the right panel, where magnetic fields of opposite polarities merge. The reconnection region example, shown in Fig.~\ref{fig3}, is located at (2,3) Mm.}
\label{fig2}
\end{figure*}

To emphasize the key features in the model, we consider the simplest case of magnetic field annihilation in a flat current sheet located at $x=0$: 
\begin{equation} 
  {\bf B} = B(x) \hat{\bf y}=(0,B(x),0) , \quad B(0)=0 . 
\end{equation}

Suppose that the merging is driven by an incompressible plasma flow of the form 
\begin{equation} 
  {\bf v} = (-v_0 \frac{x}{L}, v_0 \frac{y}{L}, 0), 
\label{linearflow}
\end{equation} 
where $L>0$ is the characteristic lengthscale and $v_0$ is the inflow speed at an outer boundary. Since the magnetic field lines are parallel to the plane of the velocity field, they are driven together by the stagnation-point flow. The resulting magnetic field build-up at the entrance to the sheet leads to a thinner sheet and a faster rate of magnetic energy dissipation. The equation of motion gives the pressure profile 
\begin{equation} 
  p(x, y) = p_0 - \frac{1}{2} B^2 - \frac{1}{2} \rho v^2.
\end{equation}
The induction equation (Eq.~\ref{inductioneq}) becomes
\begin{equation}
\eta \frac{d^2B}{dx^2} + \frac{v_0}{L} \left(x\frac{dB}{dx} + B\right) = 0,
\end{equation}
which, after integrating once, reduces to
\begin{equation} 
  \eta B' + \frac{v_0}{L} x B = E, 
\end{equation}
where $E = \eta J_0$ is the merging electric field, and $J_0 = B'(0)$ is the integration constant corresponding to the electric current density at the centre of the current sheet. The equation is integrated to yield the magnetic field profile: 
\begin{equation} 
  B_y(x) = B = J_0 l~\mbox{daw} \left(\frac{x}{l} \right) , 
\end{equation}
where 
\begin{equation}
l = \sqrt{\frac{2L\eta}{v_0}}
\label{csthickness}
\end{equation}
is the thickness of the current sheet, and $\mbox{daw}$ denotes Dawson's integral \citep[e.g.][]{Oldham-etal-2009}. 

For comparison with the simulations, analysed below, we assume the mean flow speed in the domain ($v_0  \sim 3~\mathrm{km~s^{-1}}$), the characteristic length, corresponding to the granular spatial scale ($L \sim 1~\mathrm{Mm}$), and the constant magnetic diffusivity $\eta = 2\cdot10^{10}~\mathrm{cm^2~s^{-1}}$, which is used to ensure numerical stability of the simulations. The resulting theoretical profile of the magnetic field across the current sheet is shown in Fig.~\ref{fig1_dawson}. The exact solution gives the scaling for the magnetic pile-up region and current sheet thickness, which is of the order of $30-40~\mathrm{km}$, as it is evident from the figure.

The main feature of the solution is that thinner sheets, faster inflows, and larger dissipation rates are possible when the field build-up is strong, so the magnetic field at the entrance to the current sheet is greater than $B_0$,
\begin{equation}
B_s = B\left( l \right) = J_0 l~\mathrm{daw}\left(1\right) \simeq J_0 l = B'(0)~l > B_0, 
\end{equation}
where $B_0$ is the magnetic field far from the current sheet. Such situations are likely to be the case in a high-beta photospheric plasma.

\begin{figure*}
\centering
\includegraphics{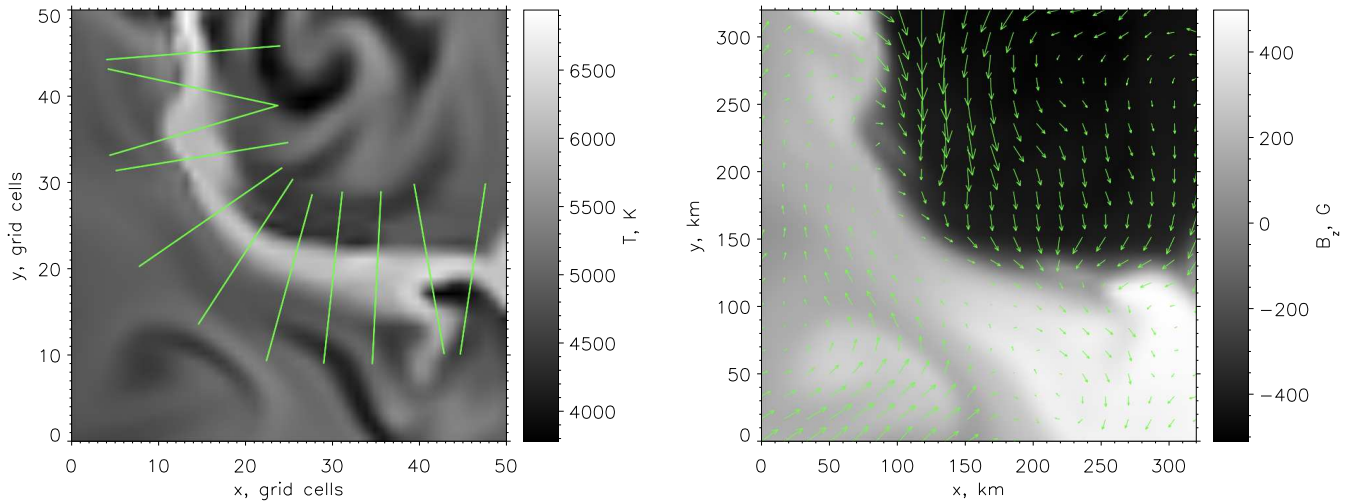}
\caption{The algorithm for determining the magnetic field and flow structure around reconnection current sheets, and magnetic field and flow structure around a simulated photospheric reconnection region. Left panel: an example of the averaging algorithm used for diagnostics of the simulated data. The background image is the horizontal cut of the temperature at the height of $300~\mathrm{km}$ above the average continuum formation height. It shows the temperature enhancements in the current sheet region due to the current dissipation. The axes units are grid cells with 6.25 km per grid cell. The green lines are constructed to align with the direction of the strongest change of vertical component of magnetic field at each point of the current sheet. The velocities are projected onto the direction parallel to the strongest gradient direction. Right panel: the vertical component of magnetic field. Arrows indicate the horizontal plasma flow direction. A flow, converging towards the reconnection region is clearly visible.}
\label{fig3}
\end{figure*}

\section{Numerical model of the solar photosphere with reconnecting magnetic fields}

We use MURaM code \citep{voegler2005} to produce the data on plasma parameters in photospheric reconnection events. The code solves the system of radiative magneto-hydrodynamic equations with constant resistivity. The numerical setup is essentially a higher-resolution version of the one used by \citet{nelson2013}, and is only briefly described here. The horizontal extent of the domain is $6 \times 6~\mathrm{Mm}$, which are resolved by $960\times960$ grid cells. The vertical extent is $1.6~\mathrm{Mm}$, resolved by 320 grid cells. This leads to the resolution of $6.25~\mathrm{km}$ and $5~\mathrm{km}$ per grid cell in the horizontal and vertical directions, respectively. The domain is positioned such that the continuum formation layer is located $1~\mathrm{Mm}$ above the bottom boundary. 4-bin non-grey radiative transport is used in the simulation. The side boundary conditions are periodic, the top boundary is closed, and the bottom boundary is open for flows. 

To simulate small-scale reconnection events in the integranular lanes of photospheric convection, we use a $4\times 4$ checkerboard pattern of vertical magnetic field with constant unsigned strength of $200~\mathrm{G}$. The magnetic field configuration is added into a well-developed non-magnetic self-consistent convection model. Then the model is evolved for 10 minutes of physical time. During these, the magnetic field, after initial uniform field annihilation phase, gets advected into the intergranular lanes of simulated granulation. The intergranular magnetic field concentrations with the (nearly vertical) field strength of $\sim 1.7~\mathrm{kG}$ randomly move along integranular lanes, being buffeted by the granular flows with the mean speed $\left< v_h\right> \approx 2.8~\mathrm{km~s^{-1}}$ at the photospheric level, occasionally come in close proximity to each other and reconnect. This leads to appearance of current sheets, which resistively dissipate and heat the plasma. Such events are demonstrated in Fig.~\ref{fig2}. In the figure, the horizontal cuts of temperature and of the vertical component of magnetic field at the height of $300~\mathrm{km}$ above the average continuum formation layer in the domain are shown. Small-scale temperature enhancements (left panel) up to $7700~\mathrm{K}$, which is $3000~\mathrm{K}$ higher than the average temperature at the corresponding height, clearly indicate the locations where the opposite polarity magnetic fields reconnect (right panel of the figure). These events are studied in the following section.

\begin{figure*}
\centering
\includegraphics{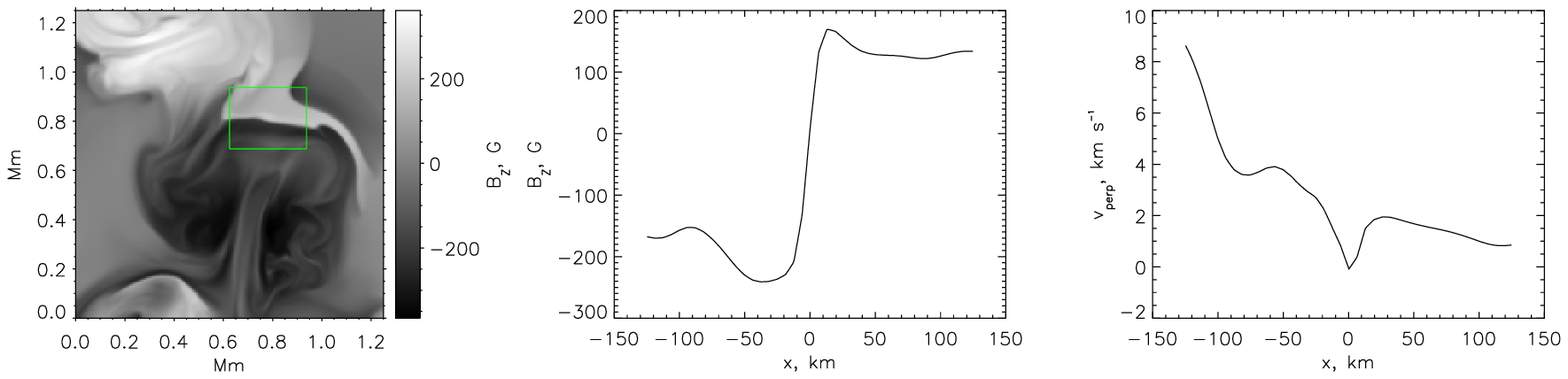}
\includegraphics{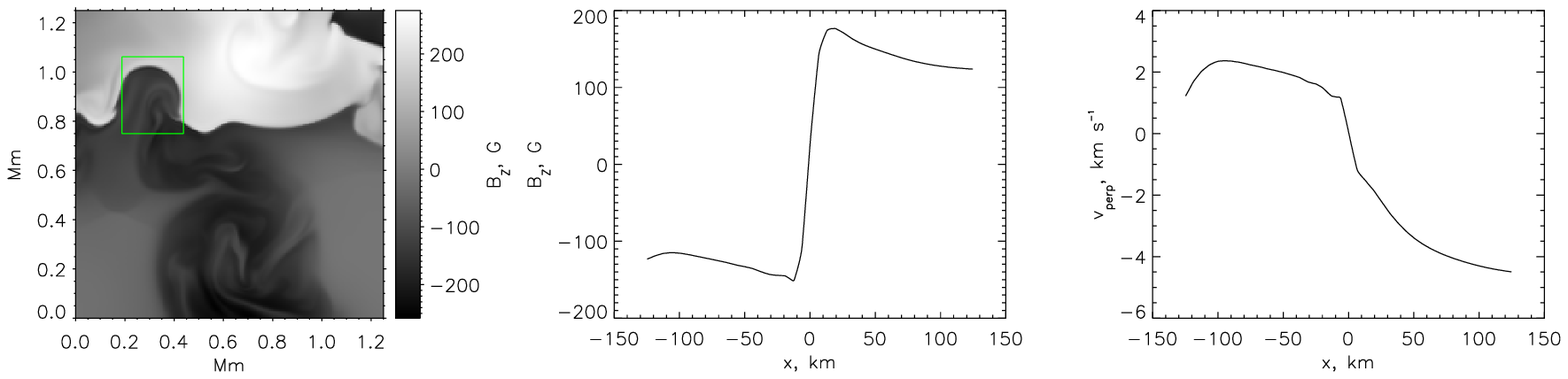}
\includegraphics{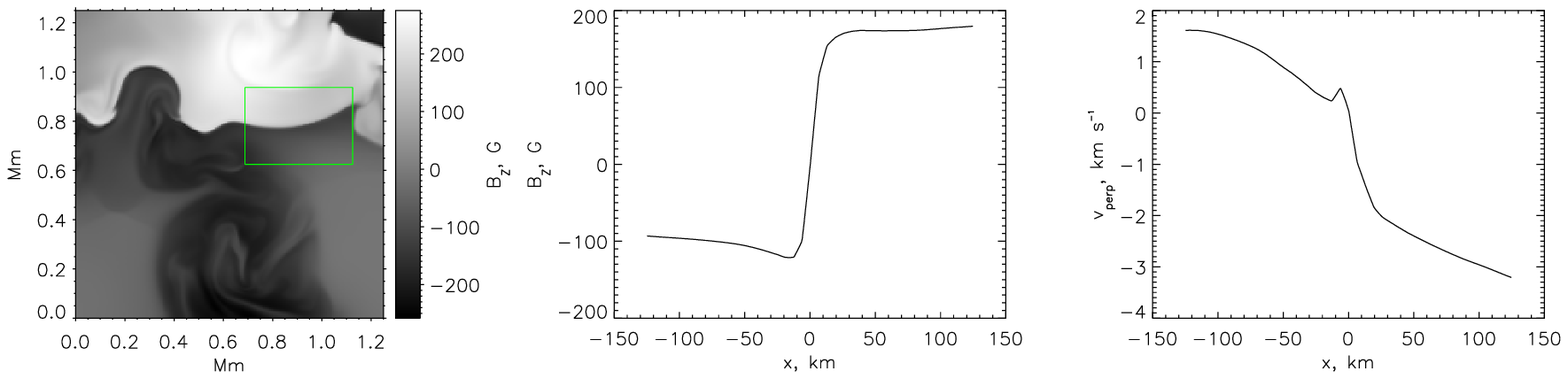}
\includegraphics{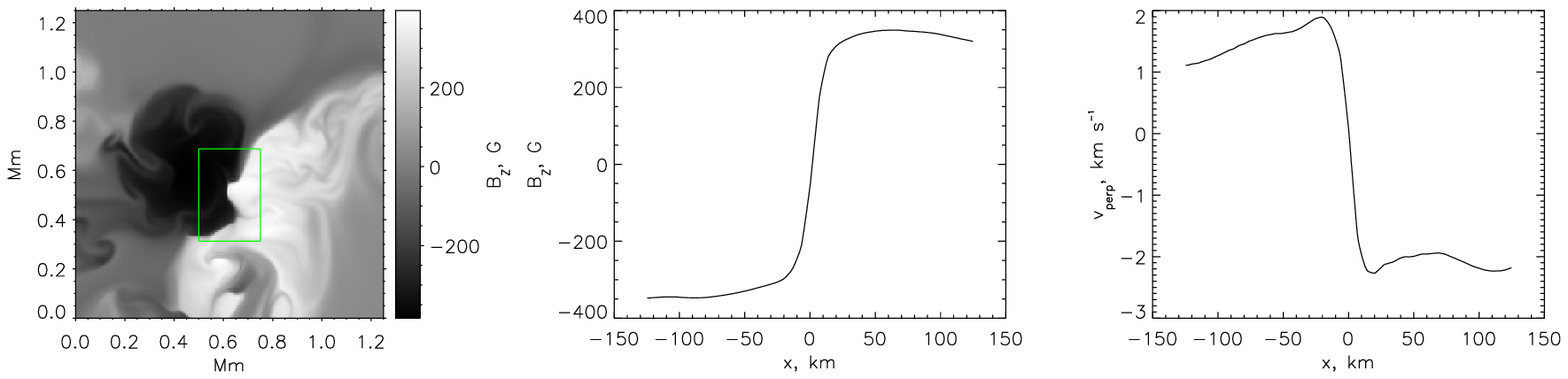}
\caption{Reconnection and magnetic flux pile-up. Left column - vertical magnetic field map around the reconnection region. Green boxes identify the averaging regions. Middle and right columns - the average vertical magnetic field and the horizontal inflow into the current sheet dependences on the distance from the current sheet, respectively. The first three rows demonstrate the magnetic flux pile-up at the scale predicted by the pile-up reconnection model. The fourth row shows stronger reconnecting magnetic fields and does not show the localised enhancements of the vertical magnetic field in proximity of the current sheet.}
\label{fig4}
\end{figure*}

\section{Simulated reconnection events}

A number of reconnection events have been identified in the simulated time series. In order to reveal the features corresponding to magnetic flux cancellation in these events, some averaging is required to remove local fluctuations of the flow and magnetic field due to turbulent convection. Therefore, to analyse the structure of magnetic field and flow surrounding the identified reconnection regions, we designed a program, which computes the average profiles of vertical magnetic field and speed of inflow into the reconnection regions. 

An example of identified reconnection region is shown in Fig.~\ref{fig3}. The temperature and vertical component of magnetic field are shown in the left and right panels of the figure, respectively. The reconnection current sheet region is, as expected, located in the region of strongest change in the magnetic field, which includes the change of the polarity. To numerically localise the reconnection current sheet, we compute the gradient of the vertical component of magnetic field $\mathbf{d} = \nabla B_z$. The maximum of the gradient therefore identifies the location of the current sheet. A unit vector in the direction of the strongest change of the vertical magnetic field $\hat{\mathbf{d}}=\mathbf{d}/d$ identifies then the direction, over which the inflow speed and the magnetic field profile are measured. The direction sign is chosen that way the gradient of the magnetic field is positive in the current sheet. The horizontal components of velocity are projected onto the direction, therefore, the positive sign of flow speed corresponds to the inflow into the current sheet from the left. Finally, the net velocity component relative to the current sheet is subtracted from the velocity profiles.

This routine is repeated in the selected region of the current sheet for each its pixel, identified as the maximum of the magnetic field gradient. Thus, the averaging is performed over the model slices, interpolated onto the direction of the strongest change of the magnetic field, and centred at the current sheet location. The result of the routine is demonstrated in the left panel of Fig.~\ref{fig3}. The green lines show the directions of averaging. The averaging is done over the distance interval $\pm 150~\mathrm{km}$ from the current sheet. 

Examples of the magnetic field and flow structure around the identified reconnections are shown in Fig.~\ref{fig4}. The examples were selected according to the following requirements. The thickness of the magnetised region has to be significantly larger than the thickness of the current sheet. Also, we aim to select the regions with smooth flows around the reconnection region. The selection is carried out by visual inspection. Finally, we aim for averaging over 50 or more automatically constructed rays.

\begin{figure*}
\centering
\includegraphics{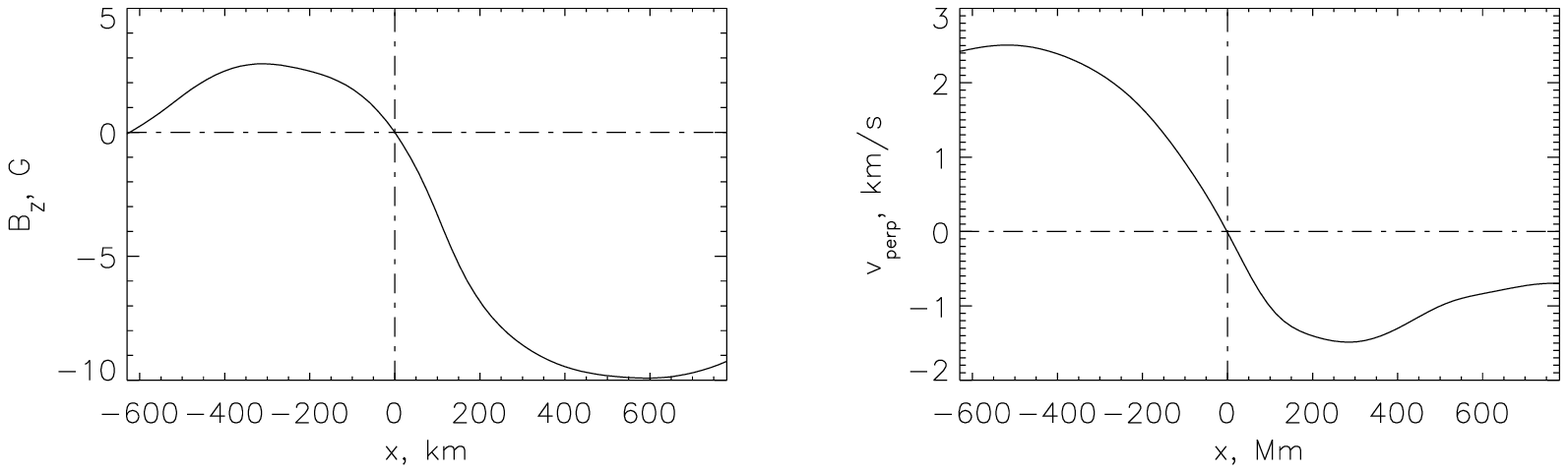}
\caption{Vertical magnetic field component (left panel) and the perpendicular velocity structure in the magnetic flux cancellation region of the cromospheric simulation.}
\label{fig5}
\end{figure*}

\begin{figure}
\centering
\includegraphics[width=\linewidth]{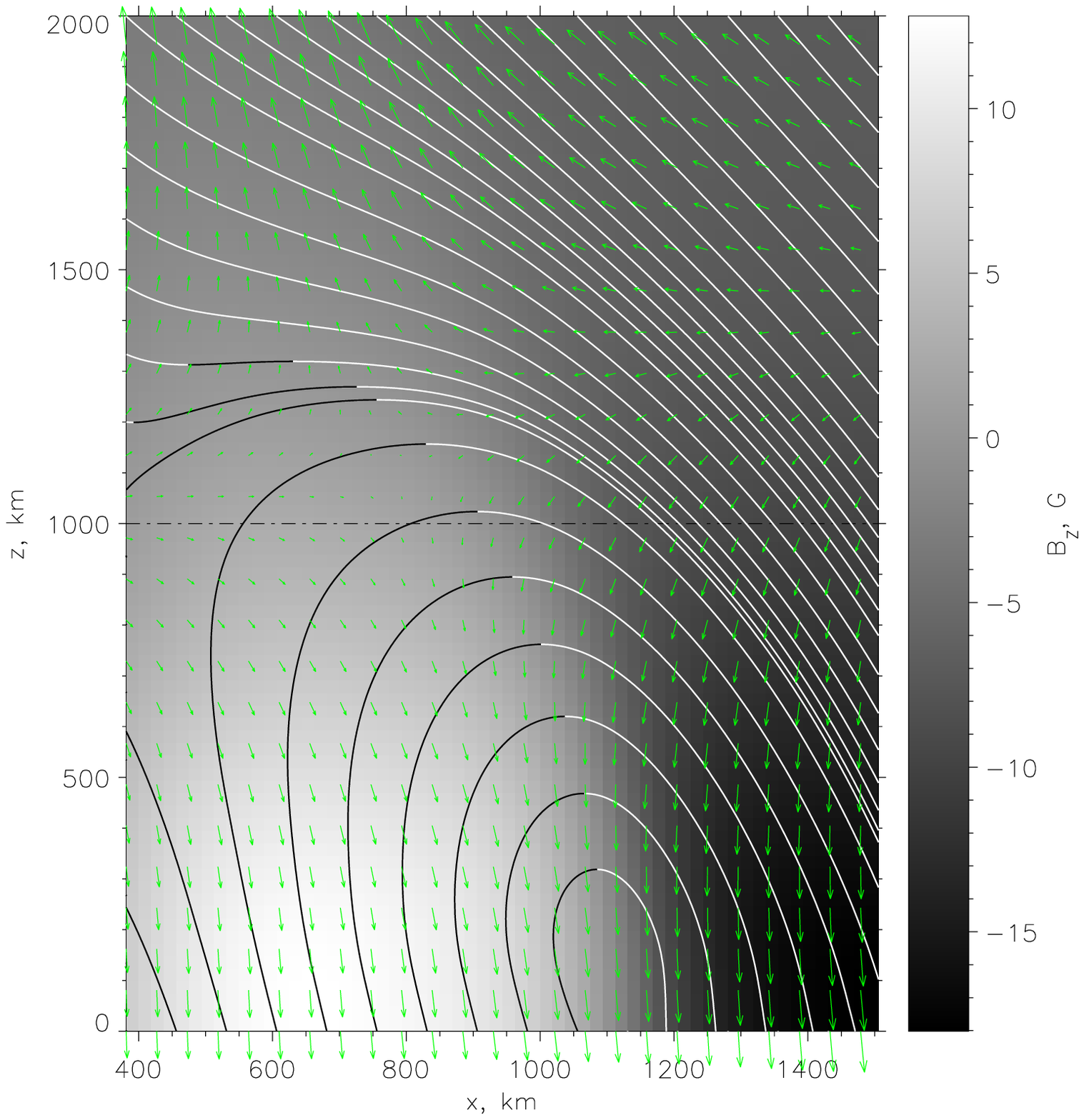}
\caption{Magnetic field geometry and velocity structure in the reconnection region. The background greyscale image shows the vertical cut ($x$ and $z$ are horizontal and vertical directions respectively) of the vertical component of the magnetic field in the domain. Height $z=0$ corresponds to the photospheric level. Solid curves show the magnetic field lines (colours correspond to the magnetic field direction). The horisontal velocity component in the plot corresponds to the velocity, perpendicular to the current sheet. The thin black dash-dotted line shows the height, where the parameters for Fig.~\ref{fig5} were measured.}
\label{fig6}
\end{figure}

Maps of the vertical component of magnetic field are shown in the left column of Fig.~\ref{fig4}. The averaging regions are identified as green boxes in the panels. The resulting average vertical magnetic field and reconnection inflow depedencies on the distance from the current sheet are shown in the second and third columns, respectively. The first three rows clearly show magnetic field pile-up around the reconnection region at $x=0$. The profile of the vertical component of magnetic field is in a good agreement with the theoretical model profile shown in Fig.~\ref{fig1_dawson}, demonstrating magnetic field intensification around the neutral line $x=0$. The distance between the local maxima is of the order of $10-20~\mathrm{km}$, again in a good agreement with the theoretical model of flux pile-up reconnection. The flows around the reconnection regions show smooth behaviour, gradually decreasing in magnitude towards the current sheet, roughly corresponding to the linear dependence of inflow as required by Eq.~\ref{linearflow}.

The fourth row of Fig.~\ref{fig4} differs significantly from the first three reconnection events. Here, the magnetic field further away from the reconnection region is significantly (almost factor of 2) stronger than in the previous cases. The magnetic field profiles do not exhibit the flux pile-up near the reconnection region, and the flow structure shows the behaviour opposite to the required by Eq.~\ref{linearflow}. Indeed, the flow velocity increases towards the reconnection region, leading to the velocity gradient of $4~\mathrm{km~s^{-1}}$, while in the cases with magnetic flux pile-up the corresponding gradient is about $2~\mathrm{km~s^{-1}}$. This indicates a different reconnection regime, more consistent with the standard Sweet-Parker reconnection with the (large due to numerical stability reasons only) constant resistivity.

Summarising the above, in resistive magneto-convection photospheric simulations with constant resistivity and opposite polarities of magnetic field, two magnetic field cancellation regimes are observed. In weaker magnetic fields, the flux pile-up regime of reconnection with a smooth inflow into the reconnection region and characteristic amplification of the magnetic field near the current sheet is identified. Therefore, the magnetic field cancellation and the magnetic energy release is intensified by the flux pile-up process. In stronger magnetic fields, contrary to the previous case, the flow profile shows a sharp jump across the reconnection region, while the magnetic field does not intensify, therefore indicating a different, Sweet-Parker-like reconnection mechanism.

\section{Chromospheric reconnection and outflows}

Similar velocity and magnetic field features were found in a simulation of ideal resistive chromospheric reconnection. The simulation is based on potential (current-free) extrapolation of the photospheric magnetic field, obtained from a low-resolution bipolar MURaM simulation \citep{2012ApJ...753L..22S, nelson2013}. The three-dimensional force-free magnetic field structure is then embedded into a hydrostatic chromospheric model and released to evolve for 60 sec of physical time. The simulation was carried out in a 3D numerical domain of the size $x \in [0,6], y \in [0,6], z \in [0,10] Mm$, covered with 240x240x400 grid cells (the effective resolution is 25 km in each direction). The constant magnetic resistivity value used in the code is $\eta=10^{12}$ cm$^{2}$ s$^{-1}$. This large resistivity value gives the thickness of the current sheet $l\approx 500~\mathrm{km}$, according to Eq.~\ref{csthickness}. Further details and simulation processes are provided in \citet{2018ApJ...856..176G}.

The simulation is analysed in a similar manner to the described in Section 3. The vertical component of the reconnecting magnetic field and the horizontal velocity, perpendicular to the manually identified current sheet, measured at the height of $1~\mathrm{Mm}$ above the photospheric level are shown in Fig.~\ref{fig5}. The constant velocity at the current sheet position was subtracted from the data. As it can be seen in the figure, the situation in this simulation is more complex. In the left part of the region around the current sheet ($x<0$), a clear pattern of magnetic pile-up (local increase of magnetic field with nearly linear inflow into the reconnection region) is present. However, in the right part of the reconnection region ($x>0$), where the magnetic field is about 3 times stronger than in the left part, no pile-up is observed, and the structure is consistent with the fourth row of Fig.~\ref{fig4} and with Sweet-Parker-like reconnection.

Fig.~\ref{fig6} shows the magnetic field geometry and the velocity structure in the vertical domain cut along the direction perpendicular to the current sheet. The simulation demonstrates a diverging velocity field, caused by compression of magnetic field lines of opposite polarities in the horizontal direction with the flow at around $1~\mathrm{Mm}$. This leads to appearance of the velocity null-point at $x=800~\mathrm{km}$ and $z=1000~\mathrm{km}$, chromospheric upflow, emerging from the reconnection region above the null-point, and a downflow below it.

\section{Conclusions}

A relatively simple, one-dimensional exact analytical description of magnetic reconnection with flux pile-up allows for magnetic flux cancellation rates, which are higher than obtained in standard Sweet-Parker-type reconnection. Flux pile-up allows to circumvent small reconnection rates of Sweet-Parker models, caused by low resistivity of the solar plasmas (although due to small ionisation fraction, magnetic Reynolds number in the chromosphere can be as low as $10-10^2$; see e.g. \citet{2012ApJ...747...87K, 2016ApJ...819L..11S}).

In this paper we used detailed three-dimensional resistive magneto-hydrodynamic models of the solar (sub-)photosphere and chromosphere with constant resistivity to demonstrate the presence of the magnetic pile-up mechanism. The photospheric data we use is generated with MURaM code and includes magnetic field concentrations of opposite polarities. The magnetic concentrations are allowed to move freely under the photospheric convective flow field and occasionally reconnect. 

Obviously, turbulent nature of three-dimensional ``realistic'' simulations does not allow appearance of local idealised solutions, directly comparable to the analytical models. Nevertheless, if appropriate averaging is applied, the simulations show that naturally generated convective flows in the simulated solar models, while being three-dimensional, produce favourable conditions for forced magnetic reconnection. The flow and magnetic field structure within the reconnection region demonstrate very good qualitative agreement with the magnetic reconnection models with flux pile-up. 

We have also demonstrated two regimes of reconnection in magneto-hydrodynamic models of the solar atmosphere. As our simulations suggest, weaker magnetic fields allow for the flux pile-up, while the stronger fields show no magnetic field intensification in proximity of the reconnection region. The latter behaviour is more consistent with the standard Sweet-Parker reconnection regime. Assuming the same magnetic field and plasma flow strength and structure, we expect this behaviour to depend on the value of magnetic diffusivity used in the simulations, with the smaller diffusivities leading to more efficient flux pile-up amplification, and higher diffusivities leading to Sweet-Parker reconnection.

\begin{acknowledgements}
The work was supported by the International Exchange programme of Royal Society (UK). VF and GV would like to thank the STFC for financial support. VF, GV, FG and JG thank Royal Society-Newton Mobility Grant NI160149, CIC-UMSNH 4.9 and CONACyT 258726. This work used the DiRAC Data Centric system at Durham University, operated by the Institute for Computational Cosmology on behalf of the STFC DiRAC HPC Facility. DiRAC is part of the National E-Infrastructure. This research was undertaken with the assistance of resources and services from the National Computational Infrastructure (NCI), which is supported by the Australian Government. JG gratefully acknowledges DGAPA postdoctoral grant to Universidad Nacional Aut\'onoma de M\'exico (UNAM).
\end{acknowledgements}

\end{document}